**Tailoring the exchange bias effect by in-plane magnetic anisotropy**

M. Ślęzak[1], T. Ślęzak[1], P. Dróżdż[1], B. Matlak[1], K. Matlak[1], J. Korecki[1,2]

[1] AGH University of Science and Technology, Faculty of Physics and Applied Computer Science, Kraków, Poland

[2] Jerzy Haber Institute of Catalysis and Surface Chemistry PAS, Kraków, Poland

ABSTRACT

We report an unusual, non-monotonous dependence of the exchange bias on the thickness of the ferromagnetic layer in a ferromagnet/antiferromagnet bilayer system. We show that in epitaxial CoO/Fe(110) bilayers, the evolution of the Fe magnetic anisotropy, which drives the thickness-induced in-plane spin-reorientation process, controls the interfacial CoO spin directions in the 0–90° range and, consequently, drastically modifies the magnitude of the hysteresis loop shift and its dependence on the thickness of the Fe layer. Our results present a new recipe for tailoring the exchange bias and antiferromagnetic spin structure by utilizing the spin-reorientation process that occurs in a ferromagnetic layer adjacent to an antiferromagnetic layer.

The superior properties of antiferromagnetic (AFM) materials, such as their resistance to external magnetic fields and negligible stray magnetic fields, promote their use in modern spintronics [1]. The application of such materials has evolved from them being static and playing a supporting role, to their being used as active and functional elements in spintronic systems. The ability to modify and detect the spin structure of AFM materials is as difficult as it is interesting from both a fundamental and an applications point of view. In low-dimensional systems, one possibility is to tailor the spin structure at a ferromagnetic/antiferromagnetic (FM/AFM) interface. This approach leads to a particularly interesting and fundamental effect, namely the exchange-bias (EB) effect [2,3] at the FM/AFM boundary, which is manifested by the shift of the hysteresis loop and an increase of coercivity. Because of the interfacial nature of the EB effect, its properties, such as anisotropy and magnitude, can be controlled by tailoring the AFM/FM interface, e.g., by controlling its chemical composition, structure, roughness, and other properties.

Another non-trivial possibility is to control the EB effect by using the intrinsic magnetic properties of the FM layer. The latter can influence the spin structure of the adjacent AFM layer and, as a result, lead to the modification of the exchange interaction and the EB effect at the FM/AFM interface. The motivation for this approach comes from a previous study [4], in which it was shown that the FM layer can govern the orientation of spins in the AFM layer at the FM/AFM interface, even though the EB effect was not reported in this study. Upon reorientation of the magnetization of the FM layer, a small canting of the interfacial spins in the AFM layer without any dire consequences on the EB effect was suggested in a previous study [5] on Fe/MnPd bilayers. Other mechanisms for controlling the EB were recently proposed, in which external factors rather than intrinsic properties were utilized [6,7]. Using properly tuned magnetic shape anisotropies in nanometer-thick lines of NiFe/FeMn bilayers, peculiar EB properties were documented [6]. In Co/CoO bilayers, the EB axis was shown to evolve from the direction imposed externally by the field-cooling procedure to the direction along the bulk AFM axis with increasing temperature [7].

In this Letter, we report on the anomalous dependence of the EB effect on the thickness of the FM layer in an FM/AFM bilayer system. We show that in epitaxial CoO/Fe(110) films grown on W(110), the well-defined and precisely controlled uniaxial magnetic anisotropy (MA) of a wedged Fe layer induces reorientation of the interfacial CoO spins. This consequently influences the magnitude of the shift of the hysteresis loop as well as its dependence on the thickness of the Fe layer. The smooth evolution of the intrinsic MA of the Fe layer is attributed to the well-known spin-reorientation-transition (SRT) process that modulates the MA strength as well as the orientation of the easy axis. Hence, both the AFM spin structure and the EB effect can be tuned by the thickness-driven SRT process in the Fe film.

To trace the evolution of the EB effect with the change in the effective in-plane MA of the FM layer, the dependence of the loop shift field $H_{EB}$ on the thickness of the Fe layer in epitaxial CoO/Fe(110) bilayers prepared on a W(110) single crystal substrate was studied. It is well known that, in Fe(110) films on W(110), the magnetization switches from the [1−10] to the [001] in-plane direction when the Fe thickness increases above a critical value [8]. This thickness-driven in-plane SRT attracted considerable attention and was also documented in a variety of bilayers, such as Au/Fe(110), Ag/Fe(110) [9,10], and, recently, in Co/Fe(110) [11]. To our knowledge, this report is the first in which Fe(110)/W(110) films coated with a metal-oxide layer (i.e., the AFM CoO) are studied.

Wedged Fe(110) films of thickness, $d_{Fe}$, ranging from 80 to 300 Å were grown on atomically clean W(110) single crystals at room temperature using molecular beam epitaxy (MBE) followed by annealing at 675 K. This produced high-quality Fe films with atomically smooth (110) surfaces. Next, a CoO adlayer 90 Å thick was grown on the Fe(110) wedge-films by reactive deposition of cobalt at a temperature of 470 K and at an $O_2$ partial pressure of $5\times10^{-7}$ mbar. The low-energy electron diffraction (LEED) pattern from the surface of the CoO-covered Fe(110)/W(110) sample indicates a hexagonal CoO(111) surface structure throughout the underlying Fe film. The details of the structural characterization are presented in supplemental material [12].

The magnetic properties of the CoO/Fe(110) system were studied *in situ*, as a function of the Fe thickness, using the longitudinal magneto-optic Kerr effect (MOKE) imaging. The magnification of the MOKE system was tuned to cover the whole sample, which was 8 mm in diameter. A series of MOKE images could be taken as a function of the external magnetic field *H,* which was applied along a chosen in-plane direction. Magnetic hysteresis loops could be extracted for any selected sample region of interest (ROI), which can be as small as one pixel. This method recently proved to be very efficient in MA studies [11] but in EB studies its advantages are even more significant. For the wedged sample, a single cooling procedure followed by the acquisition of a single MOKE "movie" as a function of the external magnetic field provided a full data set for the magnetization-reversal measurements. Moreover, all hysteresis loops were obtained under exactly the same experimental conditions, i.e., at the same sample temperature, with the same possible sample misalignments and with the same magneto-optical artifacts, if any. The size of a single ROI along the Fe-wedge gradient was 50 μm, which corresponded to an averaging of the loops over a finite Fe-thickness difference of about 2 Å.

In Figure 1, selected hysteresis loops extracted from the MOKE-imaging movie acquired at T = 183 K after cooling the sample in a remanent state, are shown for $d_{Fe}$ = 80, 110, 140 and 200 Å. From these hysteresis loops full $d_{Fe}$-dependence of the normalized magnetization in the remanence states, $M_{R1}$ and $M_{R2}$ (Fig. 2a filled symbols), coercive fields $H_{c1}$ and $H_{c2}$ (Fig. 2b filled symbols) as well as shift-field $H_{EB}$ (Fig. 2c filled symbols) as defined in [5], can be determined.

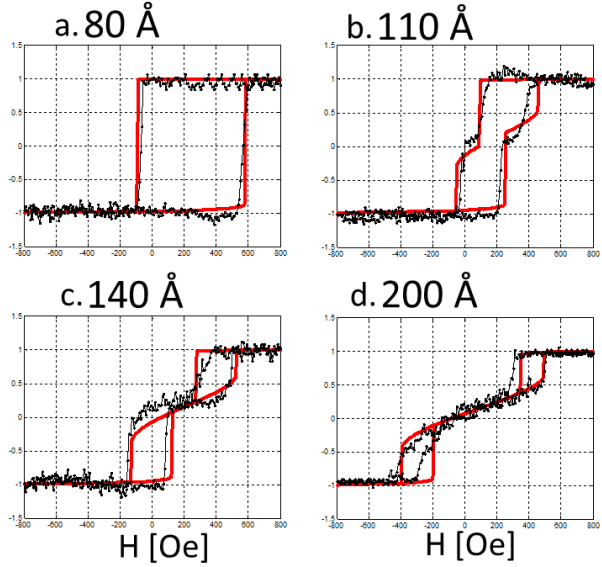

Fig. 1 Magnetic hysteresis loops measured using MOKE (at T = 183 K) and simulated for selected Fe thicknesses: (a) 80 Å, (b) 110 Å, (c) 140 Å and (d) 200 Å. The external magnetic field was applied along the [1−10] direction in the Fe(110) plane.

For $d_{Fe}$ below ~100 Å, EB-shifted square hysteresis loops, characteristic for the easy magnetization direction are observed (Fig. 1a). Therefore the values of $M_{R1}$ and $M_{R2}$ are almost constant and equal to -1 and 1, respectively (Fig. 2a). At $d_{Fe}$ = 100 Å the hysteresis loops start to evolve towards stepped loops characteristic for the hard axis of the magnetization (see Fig. 1b, c), which means that the CoO/Fe bilayer enters the SRT region. The stepped loops are also EB shifted and for this reason initially only one of the remanence states shown in Fig. 2b exhibits decreased value of the remanent magnetization, i.e. $M_{R2}$ ~ 0, while the second one is characterized by $M_{R1}$ = -1. This situation persists up to $d_{Fe}$ ~ 150 Å. At this Fe thickness the shift of the hysteresis loops is already reduced and MA enhanced enough to result in $M_{R1}$ ~ $M_{R2}$ ~ 0, as expected for typical hard axis loops in Fe on W(110) [10], similar to the one presented in Fig. 1d.

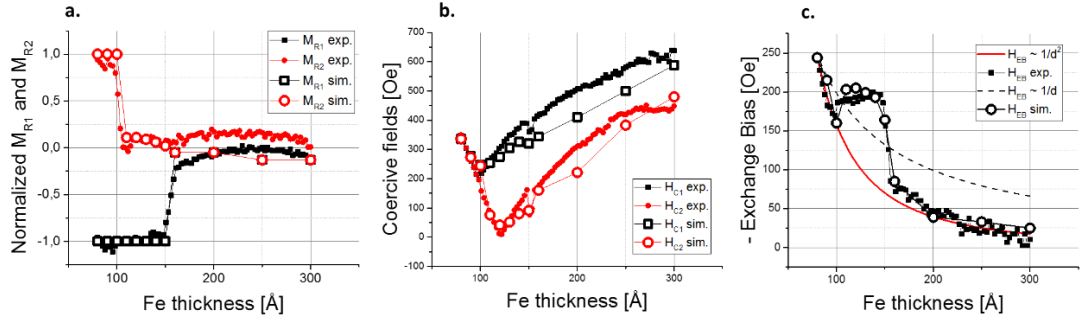

Fig. 2 Thickness dependence of MOKE results (full symbols) and simulations within coherent rotation model of magnetization reversal (open symbols) for an external magnetic field along [1-10] direction in the Fe(110) plane, at 183 K. (a) The normalized magnetization in the remanence, $M_{R1}$ and $M_{R2}$ corresponding to the states after saturation in high positive and negative external magnetic field, respectively. (b) Absolute values of coercive fields $H_{c1}$ and $H_{c2}$. (c) Shift field $H_{EB}$ compared with $1/d^2_{Fe}$ (solid line) and $1/d_{Fe}$ (dashed line) formulas.

We now focus on our main experimental finding, presented in Figure 2c, where the experimentally obtained dependence of $H_{EB}$ on $d_{Fe}$ (squares) is shown alongside the calculated $1/d_{Fe}$ (dashed line) values (which typically describe the interfacial EB effect) [13] and $1/d^2_{Fe}$ values (solid line). Two uncommon features in the $H_{EB}$ dependence on $d_{Fe}$ are to be noted. First, the dependence almost perfectly (except in the 100–150 Å $d_{Fe}$ range) varies as $1/d^2_{Fe}$, and definitely not as $1/d_{Fe}$. The $1/d^{1.9}_{FM}$ dependence of $H_{EB}$ for thick FM films was theoretically predicted [14] and $1/d^2_{FM}$ dependence was experimentally observed [15]. Such thickness dependence of the exchange bias involved the formation of planar domain walls in both AFM and FM layers. Invalidity of such modelling of our experimental results can be seen without any numerical analysis at the low thickness limit where the $1/d^2_{Fe}$ $H_{EB}$ dependence is determined from the perfectly square hysteresis loops, excluding the existence of any non-collinear Fe spin structures. Second uncommon feature in the $H_{EB}(d_{Fe})$ dependence, is the significant enhancement of $H_{EB}$ value for $d_{Fe}$ in the range 100–150 Å, with a surprising drastic drop occurring at a $d_{Fe}$ value of approximately 150 Å.

To identify the physical origin of this unusual EB behavior, we performed simulations of the magnetic hysteresis loops for the entire range of $d_{Fe}$ that was studied. The magnetization reversal is described within the coherent rotation model. Simulated magnetic hysteresis loops are obtained

from the minimization of the free enthalpy density, G, as a function of the external magnetic field, H:

$$G = - K_{EB}/d_{Fe} \cos(\Phi - \Phi_{EB}) + A \cos^2(\Phi) + B \cos^4(\Phi) - M_s H \cos(\Phi) \quad (1),$$

where $M_s$ is the saturation magnetization, $K_{EB}$ is the CoO/Fe exchange-coupling constant and $\Phi$ defines the orientation of the Fe magnetization with respect to the Fe[1−10] in-plane direction. $\Phi_{EB}$ is the exchange-bias angle, i.e., the angle between the projection of the CoO spins on the Fe(110) plane and the Fe[1−10] direction. The second- and fourth-order effective MA constants, A and B, respectively, can be defined in terms of the volume and surface MA contributions:

$$A = K_{v,p} - K_{s,p}/d_{Fe}, \quad (2)$$

$$B = K_{v,pp} - K_{s,pp}/d_{Fe}. \quad (3)$$

In Equations 2,3, $K_{v,p}$ and $K_{v,pp}$ are the second- and fourth-order volume constants of the in-plane magnetic anisotropy, while $K_{s,p}$ and $K_{s,pp}$ denote their surface analogues [10]. For all simulated hysteresis loops $K_{EB}$ was fixed at 0.5 mJ/m$^2$, which was found to provide the best fit to the experimental results. For each $d_{Fe}$, the values of A, B and $\Phi_{EB}$ were tuned to obtain the best fit between the simulated and experimental hysteresis loops. The good agreement between the black dotted (measured) and red solid (calculated) hysteresis loops in Figure 1 is to be noted. In all simulations, a 5° misalignment between the Fe[1−10] direction and the external magnetic field was assumed to account for the observed asymmetric shapes of the hysteresis loops. Results from loop modeling using Equation 1, which show perfect qualitative and good quantitative agreement with the MOKE data, are presented in Figure 2 (large open symbols). The dependencies of $M_{R1}$, $M_{R2}$, $H_{C1}$, and $H_{C2}$ with $d_{Fe}$ that were observed with MOKE are reproduced by the simulations (Figs. 2a,b). More importantly, the unusual thickness dependence of $H_{EB}$ is also perfectly reproduced using the coherent rotation model of magnetization reversal parameterized by A, B and $\Phi_{EB}$. The dependence of A and B on the inverse of the Fe thickness (1/$d_{Fe}$) and the dependence of $\Phi_{EB}$ on the Fe thickness are plotted in Figures 3a,b, respectively.

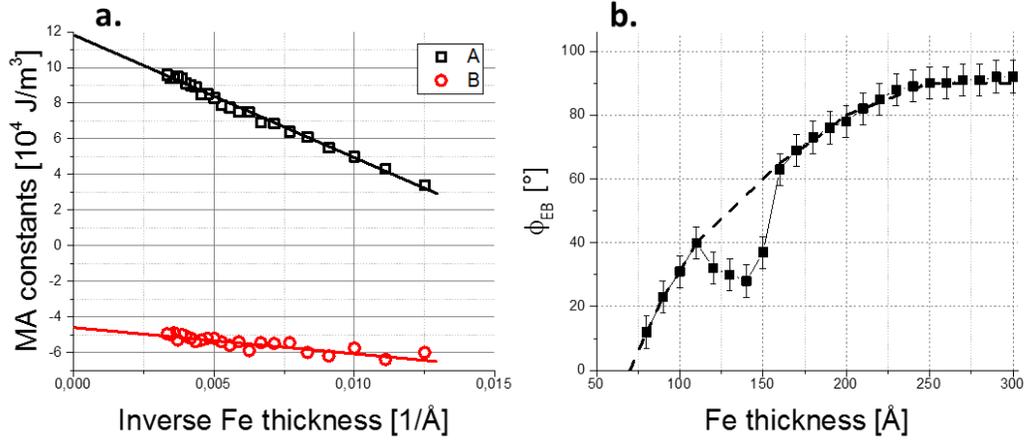

Fig. 3. a) Inverse-thickness dependence of A and B. b) Thickness dependence of $\Phi_{EB}$ with respect to the in-plane Fe[1−10] direction determined from simulations of the magnetic hysteresis loops. The dashed line in (b) represents a hypothetical [1−10] to [001] evolution of the $\Phi_{EB}$ vs Fe thickness when not perturbed in the vicinity of SRT.

As expected, A and B show a linear dependence on $1/d_{Fe}$. From linear regression, as shown in Figure 3a, it is found that $K_{v,p} = 11.8 \times 10^4$ J/m$^3$, $K_{s,p} = 0.7$ mJ/m$^2$, $K_{v,pp} = -4.6 \times 10^4$ J/m$^3$ and $K_{s,pp} = 0.1$ mJ/m$^2$.

The most important result of our study, i.e., the evolution of $\Phi_{EB}$ with $d_{Fe}$, is presented in Figure 3b. The projection of the AFM spins on the (110) plane, defined by the angle $\Phi_{EB}$, clearly reorients with increasing $d_{Fe}$. This result is interpreted as an interfacial modification of the CoO spin axis by the FM Fe layer. We conclude that Fe magnetization works as local, external magnetic field inducing magnetic anisotropy in the AFM (already during its growth) with its easy axis depending on the Fe thickness. Consequently $\Phi_{EB}$ angle is kept constant for the given hysteresis curve (as the induced AFM anisotropy is) while its thickness dependence is strongly affected by SRT process. With increasing $d_{Fe}$, the effective MA of the Fe layer evolves and its easy axis reorients from the [1−10] to the bulk-like [001] direction. Simultaneously, with this SRT in the ferromagnet, another spin reorientation takes place at the AFM/FM interface. Namely, the CoO spins rotate away from the [1−10] towards the [001] in-plane direction, as presented in Figure 3b. In the middle of this "double-SRT" the effective MA of the ferromagnet becomes zero and, therefore, the driving force for SRT in CoO vanishes. This leads, in turn, to the frustration of AFM spins at the CoO/Fe(110) interface. This frustration causes a deviation from the hypothetical [1−10] to [001] monotonous $\Phi_{EB}$ evolution, and is manifested by the deep in the dependence of $\Phi_{EB}$ on $d_{Fe}$ around 100–150 Å, as shown in Figure 3b. The SRT at the interface of the AFM CoO is therefore induced by the in-plane MA of the FM Fe layer and is reflected in the anomalous dependence of $H_{EB}$ on $d_{Fe}$ presented in Figure 2c. The monotonic parts of the $H_{EB}$ vs. $d_{Fe}$ curve follows a $1/d^2_{Fe}$ behavior instead of the expected $1/d_{Fe}$ behavior, because of the superposition of two effects: the interfacial nature of the EB effect and the continuously rotating EB axis defined by $\Phi_{EB}$. On the other hand, the intriguing plateau followed by a drastic drop of $H_{EB}(d_{Fe})$ dependence in 100 – 150 Å range (Fig. 2c) results from the non-monotonous $\Phi_{EB}(d_{Fe})$ dependence seen in Fig. 3b. We highlight the possibility to simultaneously control the interfacial spin axis of an antiferromagnet and to tailor the EB effect in FM/AFM systems with the well-defined MA of a ferromagnet. Moreover, reported magnetic anisotropy of the AFM that is induced by a FM layer, well explains the presence of a spontaneous EB in many FM/AFM systems. This kind of MA engineering could be used in future experiments with both self-organized and patterned nanostructures. One such promising example is that of Fe islands on W(110). In well-prepared Fe/W samples, [1−10] as well as [001] magnetized nanostructures can exist. In addition, another possibility, i.e., nanostructures with coexisting [1−10] and [001] magnetic domains, has also been reported [16]. When such nanostructures are capped with CoO, a strong local sensitivity of the EB effect on the magnetic state of a given Fe island can be expected.

This work was supported by the National Science Center Poland (NCN) under Project No. 2011/02/A/ST3/00150. The research was performed in the framework of the Marian Smoluchowski Krakow Research Consortium, the Leading National Research Centre (KNOW), which is supported by the Polish Ministry of Science and Higher Education.